\documentclass[aps,prb,twocolumn,showpacs,groupedaddress,preprintnumbers,amsmath,amssymb]{revtex4}

\usepackage {amsmath,amssymb,epsfig,color}
\usepackage {graphicx}
\usepackage {dcolumn}
\usepackage {bm}

\definecolor{darkgreen}{rgb}{0,0.5,0}
\definecolor{darkred}{rgb}{0.5,0,0}

\begin{document}

\title{Doping-dependent evolution of low-energy excitations and quantum phase transitions within effective model for High-$T_c$ copper oxides}

\author {M.M.~Korshunov $^{1,2}$}
 \email {maxim@mpipks-dresden.mpg.de}
\author {S.G.~Ovchinnikov $^1$}
 \affiliation {$^1$ L.V. Kirensky Institute of Physics, Siberian Branch of Russian Academy of Sciences, 660036 Krasnoyarsk, Russia}
 \affiliation {$^2$ Max-Planck-Institut f\"{u}r Physik komplexer Systeme, D-01187 Dresden, Germany}

\date{\today}

\begin{abstract}
In this paper a mean-field theory for the spin-liquid paramagnetic non-superconducting phase 
of the p- and n-type High-$T_c$ cuprates is developed.
This theory applied to the effective $t-t'-t''-J^*$ model with the {\it ab initio} 
calculated parameters and with the three-site correlated hoppings. 
The static spin-spin and kinematic correlation functions beyond Hubbard-I 
approximation are calculated self-consistently. 
The evolution of the Fermi surface and band dispersion 
is obtained for the wide range of doping concentrations $x$.
For p-type systems the three different types of behavior are found and the 
transitions between these types are accompanied by the changes in the Fermi surface topology. 
Thus a quantum phase transitions take place at $x=0.15$ and at $x=0.23$.
Due to the different Fermi surface topology we found for n-type cuprates 
only one quantum critical concentration, $x=0.2$.
The calculated doping dependence of the nodal Fermi velocity and the effective mass 
are in good agreement with the experimental data.
\end{abstract}

\pacs{74.72.-h; 74.25.Jb; 73.43.Nq; 71.18.+y}

\maketitle

\section{Introduction \label{section:intro}}

Discovered almost 20 years ago, High-$T_c$ copper oxides still remain a challenge of the modern condensed matter physics.
It is not only due to unconventional superconducting state with a highest superconducting transition temperature $T_c$ 
ever observed. Also they reveal evolution from an undoped antiferromagnetic (AFM) insulator to 
an almost conventional, though highly correlated \cite{nakamae2005}, Fermi liquid system at the overdoped side of the phase diagram. 
Between these two regimes the system exhibits strongly correlated, or so-called ``pseudogap'' metallic behavior 
up to an optimal doping concentration $x_{opt} \approx 0.16$.

Recent significant improvements of experimental techniques, especially of the Angle-Resolved Photoemission Spectroscopy (ARPES) and the Scanning Tunneling Microscopy (STM), revealed many exciting features of this doping dependent evolution. 
First of all, the Fermi surface (FS) at low doping concentrations $x$ has been measured \cite{shen2005}.
Together with the previous measurements on optimally and overdoped samples 
(see e.g. review \cite{damascelli2003} and references therein), 
these observations provide a unified picture of the doping dependent FS, 
which changes from the ``Fermi arcs'' \cite{yoshida2003} in the underdoped compounds 
to the ``large'' Fermi surface in the overdoped systems.
Though this change is smooth, it occurs around the optimal doping concentration.
Also, the observed evolution is consistent with the Hall coefficient $R_H$ measurements \cite{ando2004RH}.

The pseudogap behavior observed in ARPES is also tracked in the transport measurements. 
In particular, the resistivity curvature mapping over $T-x$ phase diagram clearly 
demonstrates crossover between underdoped and overdoped regimes \cite{ando2004}. 
And the in-plane resistivity $\rho_{ab}$ shows $T$-linear dependence only around $x_{opt}$.

The drastic change of the quasiparticle dynamics around the optimal doping was found 
by the time-resolved measurements of the photoinduced change in reflectivity for 
Bi$_2$Sr$_2$Ca$_{1-y}$Dy$_y$Cu$_2$O$_{8+\delta}$ (BSCCO) \cite{gedik2005}.
Namely, the spectral weight shifts expected for a BCS superconductor
can account for the photoinduced response in overdoped, but not underdoped BSCCO.
This agrees with the observed difference of the low-frequency spectral weight transfer 
in normal and superconducting states on under- and over-doped samples \cite{bontemps2004,carbone2006}.

Meanwhile, the integral characteristics of the system demonstrate more smooth behavior upon increase of the doping $x$. 
The dependence of the chemical potential shift $\Delta \mu$ on $x$ shows pinning at $x<x_{opt}$, and evolves smoothly at higher doping concentrations \cite{harima2001}. The measured nodal Fermi velocity $v_F$ is almost doping-independent 
within experimental error of 20\% \cite{zhou2003,kordyuk2005}.
The experiments involving combination of dc transport and infrared spectroscopy revealed 
an almost constant effective electron mass $m^*/m = 3.8 \pm 2$ in the underdoped and 
slightly overdoped La$_{2-x}$Sr$_x$CuO$_4$ (LSCO) and YBa$_2$Cu$_6$O$_y$ (YBCO) \cite{padilla2005}.
Low-$x$ effective mass dependence contradicts predictions of the 
Brinkman-Rice metal-insulator transition theory \cite{brinkman1970}, 
predicting divergence of the $m^*$ at the point of transition.
One of the main drawbacks in this theory is that the magnetic 
correlations were neglected. Thus the discrepancy with the experiment emphasizes 
the importance of these correlations in High-$T_c$ copper oxides.

From the theoretical point of view, the description of the crossover between 
almost localized picture and the Fermi liquid regime is very difficult. 
Starting from the Fermi liquid approach one may discuss the overdoped 
and, partly, optimally doped region, while for underdoped and undoped samples
this approach is not applicable.
The strong-coupling Gutzwiller approximation \cite{gutzwiller1963} for the Hubbard model 
provides a good description for the correlated metallic system. This approximation is 
equivalent \cite{gebhard1990,gebhard1991} to the mean-field saddle-point solution 
within a slave-boson approach \cite{kotliar1986}. At the same time, as shown 
within $1/d$ expansion \cite{gebhard1990}, with $d$ being the dimensionality of 
the lattice, the Gutzwiller approximation is equivalent to the case of $d=\infty$.
Obviously, for the quasi-two-dimensional systems such as High-$T_c$ copper oxides 
this is not a proper limit. The same applies to the Dynamical Mean-Field Theory 
(DMFT) \cite{metzner1989,georges1996}, which is exact only for $d=\infty$. 
In this limit the short-range magnetic fluctuations are excluded. 
It is not a good starting point for the system with long-range AFM order 
at low $x$ and short-range AFM correlations in the underdoped region.

To describe the doping dependent evolution of the low-energy excitations 
we develop a strong-coupling mean-field theory for the paramagnetic 
non-superconductive phase of High-$T_c$ copper oxides starting from the local limit. 
To go beyond the usual Hubbard-I approximation a self-consistently calculated 
static spin-spin and kinematic correlation functions are taken into account.
Within this approximation in the framework of the effective $t-t'-t''-J^*$ model with {\it ab initio} calculated 
parameters we obtain a doping-dependent evolution of the FS, effective mass and nodal Fermi velocity.
The analysis of the low-energy excitations behavior for p-type cuprates
yields two quantum phase transitions associated with the change of the FS topology.
For n-type cuprates we observe only single quantum critical concentration.
The key aspect in these findings is an adequate description of the 
electron scattering by the short range magnetic fluctuations
accompanied by the three-site correlated hoppings.

The paper is organized as follows. In Section~\ref{section:model} 
the effective model and the approximations are described.
The results of the calculations for p-type cuprates are presented in 
Section~\ref{section:ptype}. Also, the critical comparison of the 
$t-t'-t''-J^*$ and $t-t'-t''-J$ models is made, and the role
of short range magnetic order is discussed.
Section~\ref{section:ntype} contains results for n-type cuprates.
The last Section summarize this study, and the main points are discussed.

\section{Model and Approximations \label{section:model}}

High-$T_c$ cuprates belong to a class of strongly correlated systems where 
the standard local density approximation (LDA) schemes and the weak-coupling perturbation 
theories yield an inappropriate results. To overcome this difficulty recently 
we have developed an LDA+GTB method \cite{korshunov2005}. In this method 
the {\it ab initio} LDA calculation is used to construct the Wannier functions 
and to obtain the single electron and Coulomb parameters of the multiband 
Hubbard-type model. Within this model the electronic structure 
in the strong correlation regime is calculated by the Generalized Tight-Binding (GTB) 
method \cite{ovchinnikov1989,gavrichkov2000}. The latter combines the exact diagonalization of 
the model Hamiltonian for a small cluster (unit cell) with perturbative treatment 
of the intercluster hopping and interactions. For undoped and weakly doped LSCO and 
Nd$_{2-x}$Ce$_x$CuO$_4$ (NCCO) this scheme results in a charge transfer insulator 
with a correct value of the gap and the dispersion of bands in agreement with 
the experimental ARPES data (see Ref.~\cite{korshunov2005} for details).

Then this multiband Hubbard-type Hamiltonian was mapped 
onto an effective low-energy model \cite{korshunov2005}.
Parameters of this effective model were obtained directly 
from the {\it ab initio} parameters of the multiband model.
The low-energy model appears to be the $t-t'-t''-J^*$ model 
($t-t'-t''-J$ model with the three-site correlated hoppings) for n-type cuprates 
and the {\it singlet-triplet} $t-t'-t''-J^*$ model for p-type systems. 
However, for $x<0.7$ in a phase without a long-range magnetic order the role of the triplet state 
and the singlet-triplet hybridization is negligible \cite{korshunov2006}. 
Therefore, the triplet state could be omitted and in the present paper we will 
describe the low-energy excitations in the single-layer p- and n-type cuprates 
within the $t-t'-t''-J^*$ model.

To write down the model Hamiltonian we use the Hubbard $X$-operators \cite{hubbard1964}:
$X_f^\alpha \leftrightarrow X_f^{n,n'} \equiv \left| n \right> \left< n' \right|$.
Here index $\alpha \leftrightarrow (n,n')$ enumerates quasiparticle with energy 
$\omega_\alpha = \varepsilon_n (N + 1) - \varepsilon_{n'} (N)$,
where $\varepsilon_n$ is the $n$-th energy level of the $N$-electron system.
The commutation relations between X-operators are quite complicated, 
i.e. two operators commute on another operator, not a $c$-number.
Nevertheless, depending on the difference of the number of fermions in states $n$ and $n'$
it is possible to define quasi-Fermi and quasi-Bose type operators in terms of obeyed statistics.
In this notations the Hamiltonian of the $t-t'-t''-J^*$ model in the hole representation have the form:
\begin{eqnarray}
\label{eq:H}
H&=&\sum_{f, \sigma} \left( \varepsilon_0 -\mu \right) X_{f}^{\sigma, \sigma} + 
\sum_{f \neq g, \sigma} t_{fg} X_{f}^{\sigma, 0} X_{g}^{0, \sigma} \nonumber \\
&+& \sum_{f \neq g} J_{fg} \left( \vec{S}_f \vec{S}_g - \frac{1}{4} n_f n_g \right) + H_3.
\end{eqnarray}
Here $\mu$ is the chemical potential, $\vec{S}_f$ is the spin operator,
$S_f^+=X_f^{\sigma, \bar \sigma}$, $S_f^-=X_f^{\bar \sigma, \sigma}$, 
$S_f^z=\frac{1}{2} \left( X_f^{\sigma, \sigma}-X_f^{\bar \sigma, \bar \sigma} \right)$,
$n_f=\sum \limits_{\sigma} X_f^{\sigma, \sigma}$ is the number of particles operator, 
$J_{fg} = 2 \tilde{t}_{fg}^{2} / E_{ct}$ is the exchange parameter, 
$E_{ct}$ is the charge-transfer gap.
In the notations of Ref.~\cite{korshunov2005} the hopping matrix elements $t_{fg}$ 
corresponds to $t_{fg}^{SS}$ and $-t_{fg}^{00}$ for p- and n-type cuprates, respectively, 
and $\tilde{t}_{fg} = t_{fg}^{0S}$.
Hamiltonian $H_3$ contains the three-site interaction terms:
\begin{equation}
\label{eq:H_3}
H_3 = \sum\limits_{f \neq g \neq m, \sigma} \frac{\tilde{t}_{fm} \tilde{t}_{mg}}{E_{ct}} 
\left( X_f^{\sigma 0} X_m^{\bar \sigma \sigma} X_g^{0\bar \sigma} 
- X_f^{\sigma 0} X_m^{\bar \sigma \bar \sigma} X_g^{0\sigma} \right).
\end{equation}

There is a simple correspondence between $X$-operators and single-electron 
annihilation operators:
$a_{f \lambda \sigma} = \sum\limits_\alpha \gamma_{\lambda \sigma}(\alpha) X_f^\alpha$, 
where the coefficients $\gamma_{\lambda \sigma}(\alpha)$ determines the partial weight of the 
quasiparticle $\alpha$ with spin $\sigma$ and orbital index $\lambda$.
These coefficients are calculated straightforwardly within the GTB scheme.
In the considered case there is only one quasi-Fermi-type quasiparticle, $\alpha=(0,\sigma)$, 
with $\gamma_{\lambda \sigma}(\alpha)=1$, 
and the Hamiltonian in the generalized form in momentum representation is given by:
\begin{eqnarray}
H&=&\sum\limits_{\vec{k}, \sigma} \left( \varepsilon_0 - \mu \right) X^{\sigma,\sigma}_{\vec{k}}
+ \sum\limits_{\vec{k}} \sum\limits_{\alpha,\beta} t_{\vec{k}}^{\alpha \beta} {X_{\vec{k}}^\alpha}^\dag X_{\vec{k}}^{\beta} \nonumber \\
&+& \sum\limits_{\vec{p}, \vec{q}} \sum\limits_{\alpha,\beta,\sigma,\sigma'} V_{\vec{p} \vec{q}}^{\alpha \beta, \sigma \sigma'} {X_{\vec{p}}^\alpha}^\dag X_{\vec{p}-\vec{q}}^{\sigma,\sigma'} X_{\vec{q}}^{\beta}.
\end{eqnarray}

The Fourier transform of the two-time retarded Green function in the energy representation, 
$G_{\lambda}(\vec{k},E) = \left<\left< a_{\vec{k} \lambda \sigma} \left| a_{\vec{k} \lambda \sigma}^\dag \right. \right>\right>_E$, 
can be rewritten in terms of the matrix Green function 
$\left[ \hat{D}(\vec{k},E) \right]_{\alpha \beta} = \left<\left< X_{\vec{k}}^\alpha \left| {X_{\vec{k}}^\beta}^\dag \right. \right>\right>_E$:
\begin{equation}
G_{\lambda}(\vec{k},E) = \sum\limits_{\alpha,\beta} \gamma_{\lambda \sigma}(\alpha) \gamma_{\lambda \sigma}^*(\beta) 
D^{\alpha \beta}({\vec{k}},E).
\end{equation}

The diagram technique for Hubbard X-operators has been developed \cite{zaitsev1975,izumov1991} 
and the generalized Dyson equation \cite{ovchinnikov_book2004} 
in the paramagnetic phase ($\left< X_{0}^{\sigma,\sigma}\right>=\left< X_{0}^{\bar \sigma,\bar \sigma}\right>$) 
reads:
\begin{eqnarray} \label{eq:D}
\hat{D}(\vec{k},E) &=& \left[ \hat{G}_0^{-1}(E) 
- \hat{P}(\vec{k},E) \hat{t}_{\vec{k}} 
- \hat{P}(\vec{k},E) \hat{V}_{\vec{k}\vec{k}}^{\sigma,\sigma} \left< X_{0}^{\sigma,\sigma}\right> \right. \nonumber\\
&+& \left. \hat{\Sigma}(\vec{k},E) \right]^{-1} \hat{P}(\vec{k},E).
\end{eqnarray}

Here, $\hat{G}_0^{-1}(E)$ is the exact local Green function, 
$G_0^{\alpha \beta}(E) = \delta_{\alpha \beta} / \left[ {E - \left({\varepsilon_n - \varepsilon_{n'} } \right)} \right]$, 
$\hat{\Sigma}(\vec{k},E)$ and $\hat{P}(\vec{k},E)$ are the self-energy and the strength operators, respectively. 
The presence of the strength operator is due to the redistribution of the spectral 
weight between the Hubbard subbands, that is an intrinsic feature of the strongly correlated electron 
systems. It also should be stressed that $\hat{\Sigma}(\vec{k},E)$ in Eq.~(\ref{eq:D}) is the 
self-energy in $X$-operators representation and therefore it differs from the 
self-energy entering Dyson equation for the weak coupling perturbation theory for $G_{\lambda}(\vec{k},E)$.

Within Hubbard-I approximation \cite{hubbard1963} the self-energy $\hat{\Sigma}(\vec{k},E)$ 
is equal to zero and the strength operator $\hat{P}(\vec{k},E)$ is replaced by 
$P^{\alpha \beta}(\vec{k},E) \to P^{\alpha \beta} = \delta_{\alpha \beta} F_\alpha $, where 
$F_{\alpha(n,n')} = \left< X_f^{n,n} \right> + \left< X_f^{n',n'} \right>$ is the occupation factor. 

Taking into account that in the considered paramagnetic phase 
$\left< X_f^{\sigma,\sigma} \right> = \frac{1-x}{2}$,
$\left< X_f^{0,0} \right> = x$, 
with $x$ being the doping concentration,
after all substitutions 
and treating all $\vec{k}$-independent terms as the chemical potential renormalization,
the generalized Dyson equation for the Hamiltonian~(\ref{eq:H}) becomes:
\begin{eqnarray}
\label{eq:D_H}
D(\vec{k},E) &=& \left[ E - (\varepsilon_0 - \mu) 
- \frac{1+x}{2} t_{\vec{k}} \right. \nonumber\\
&-& \left. \frac{1+x}{2} \frac{\tilde{t}_{\vec{k}}^2}{E_{ct}} \frac{1-x}{2}
+ \Sigma(\vec{k},E) \right]^{-1} \frac{1+x}{2}.
\end{eqnarray}

To go beyond the Hubbard-I approximation we have to calculate $\Sigma(\vec{k},E)$.
For this purpose we use an equations of motion method for the $X$-operators \cite{plakida1989}.
The exact equation of motion for $X_k^{\alpha}$ is:
\begin{eqnarray}
i \dot X_{\vec{k}}^{\alpha} &=& \left[ X_{\vec{k}}^{\alpha}, H \right] 
= (\varepsilon_0-\mu) X_{\vec{k}}^{\alpha} + L_{\vec{k}}^{\alpha} \nonumber\\
&=& (\varepsilon_0-\mu) X_{\vec{k}}^{\alpha} + {L_{\vec{k}}^{\alpha}}^{(0)} 
+ ( L_{\vec{k}}^{\alpha} - {L_{\vec{k}}^{\alpha}}^{(0)} ).
\end{eqnarray}
Here ${L_{\vec{k}}^{\alpha}}^{(0)}$ is the linearized and decoupled in Hubbard-I approximation operator $L_{\vec{k}}^{\alpha}$,
\begin{equation}
{L_{\vec{k}}^{\alpha}}^{(0)}= \frac{1+x}{2} t_{\vec{k}} + \frac{1+x}{2} \frac{\tilde{t}_{\vec{k}}^2}{E_{ct}} \frac{1-x}{2}.
\end{equation}
Let us define $\tilde{L}_{\vec{k}}^{\alpha} = L_{\vec{k}}^{\alpha} - {L_{\vec{k}}^{\alpha}}^{(0)}$ 
and linearize it with respect to $X_{\vec{k}}^{\alpha}$:
\begin{eqnarray}
\tilde{L}_{\vec{k}}^{\alpha} = \sum\limits_{\beta} T_{\vec{k}}^{\alpha \beta} X_{\vec{k}}^{\beta} 
+ \tilde{L}_{\vec{k}}^{\alpha (irr)},
\end{eqnarray}
where $T_{\vec{k}}^{\alpha \beta} = \frac{\left< \left\{ \tilde{L}_{\vec{k}}^{\alpha},{X_{\vec{k}}^{\beta}}^\dag \right\} \right>}
{\left< \left\{ X_{\vec{k}}^{\beta},{X_{\vec{k}}^{\beta}}^\dag \right\} \right>}$
are the coefficients of the linearization.
All effects of the finite quasiparticle lifetime are contained in the irreducible part 
$\tilde{L}_{\vec{k}}^{\alpha (irr)}$. In this paper we neglect it, $\tilde{L}_{\vec{k}}^{\alpha (irr)} \to 0 $.

Since the exact equation for the Green function is given by  
$ E \left<\left< X_{\vec{k}}^\alpha \left| {X_{\vec{k}}^\beta}^\dag \right. \right>\right>_E 
= \left<\left\{ X_{\vec{k}}^\alpha, {X_{\vec{k}}^\beta}^\dag \right\}\right> 
+ \left<\left< i \dot X_{\vec{k}}^{\alpha} \left| {X_{\vec{k}}^\beta}^\dag \right. \right>\right>_E$, 
it is straightforward to find that in our approximation $T_{\vec{k}}^{\alpha \beta}$ corresponds to the self-energy:
\begin{eqnarray}
\label{eq:Tapprox}
\hat{\Sigma}(\vec{k},E) = - \hat{T}_{\vec{k}}.
\end{eqnarray}

Introducing notations for the static spin-spin correlation functions 
\begin{equation}
\label{eq:C}
C_{\vec{q}} = \sum\limits_{f,g} e^{-i(f-g)\vec{q}} \left< X_f^{\sigma \bar\sigma} X_g^{\bar\sigma \sigma} \right> = 2 \sum\limits_{\vec{r}} e^{-i \vec{r} \vec{q}} \left< S_{\vec{r}}^z S_0^z \right>,
\end{equation} 
and for the kinematic correlation functions 
\begin{equation}
\label{eq:K}
K_{\vec{q}} = \sum\limits_{f,g} e^{-i(f-g){\vec{q}}} \left< X_f^{\sigma 0} X_g^{0\sigma} \right>,
\end{equation}
the expression for the quasiparticle self-energy becomes:
\begin{widetext}
\begin{eqnarray}
\label{eq:Sigma}
\Sigma(\vec{k},E) &=& \frac{2}{1+x} \frac{1}{N} \sum\limits_{\vec{q}} 
  \left[ t_{\vec{q}} 
  - \frac{1-x}{2} J_{\vec{k}-\vec{q}} 
  - x \frac{\tilde{t}_{\vec{q}}^2}{E_{ct}} 
  - \frac{1+x}{2} \frac{2 \tilde{t}_{\vec{k}} \tilde{t}_{\vec{q}}}{E_{ct}} \right] K_{\vec{q}} \nonumber\\
  &-& \frac{2}{1+x} \frac{1}{N} \sum\limits_{\vec{q}} 
  \left[ t_{\vec{k}-\vec{q}} 
  - \frac{1-x}{2} \left( J_{\vec{q}} - \frac{\tilde{t}_{\vec{k}-\vec{q}}^2}{E_{ct}} \right)
  - \frac{1+x}{2} \frac{2 \tilde{t}_k \tilde{t}_{\vec{k}-\vec{q}}}{E_{ct}} 
  \right] \frac{3}{2} C_{\vec{q}}.
\end{eqnarray}
\end{widetext}
Here $N$ is the number of vectors in momentum space.

Until now we have made two major approximations. 
First, we neglected irreducible part of the self-energy, $\tilde{L}_{\vec{k}}^{\alpha (irr)}$,
thus allowing quasiparticles to have infinite lifetime.
Second, and this is not so obvious, in Eq.~(\ref{eq:D_H}) for 
the strength operator $\hat{P}(\vec{k},E)$ we drop out 
corrections beyond Hubbard-I approximation. 
These dropped corrections also lead to the finite quasiparticle lifetime.
The consequence of these approximations will be discussed later.

Kinematic correlation functions (\ref{eq:K}) are calculated straightforwardly with the help of Green function~(\ref{eq:D_H}).
The spin-spin correlation functions for the $t-J$ model with three-site correlated hoppings $H_3$ 
were calculated in Ref.~\cite{valkov2005}. 
In this paper the equations of motion for the spin-spin Green function 
$\left<\left< X_f^{\sigma \bar\sigma} \left| X_g^{\bar\sigma \sigma} \right. \right>\right>_\omega$ 
were decoupled in the rotationally invariant quantum spin liquid phase, 
similar to Refs.~\cite{shimahara1991,barabanov1992}.
The results of this approach for a static magnetic susceptibility for the $t-J$ model 
are similar to those obtained by other methods~\cite{sherman2002,vladimirov2005}.
Higher-order correlation functions appearing due to the $H_3$ term are decoupled in the following way:
\begin{eqnarray*}
\left<\left< X_m^{\bar\sigma \bar\sigma} X_n^{\sigma 0} X_l^{0\bar \sigma} \left| \right. X_j^{\bar\sigma \sigma} \right>\right>_\omega \to \left< X_m^{\bar\sigma \bar\sigma} \right> \left<\left< X_n^{\sigma 0} X_l^{0\bar \sigma} \left| \right. X_j^{\bar\sigma \sigma} \right>\right>_\omega \\
\left<\left< X_m^{\sigma \bar\sigma} X_l^{\sigma 0} X_n^{0\sigma} \left| \right. X_j^{\bar\sigma \sigma} \right>\right>_\omega \to \left< X_l^{\sigma 0} X_n^{0\sigma} \right> \left<\left< X_m^{\sigma \bar\sigma} \left| \right. X_j^{\bar\sigma \sigma} \right>\right>_\omega
\end{eqnarray*}
Thus, higher-order kinematic and spin-spin scattering channels are decoupled.

After the terms proportional to $x \left(2 \tilde{t}_{01}/ E_{ct} \right)^2$ being neglected, 
the expression for Fourier transform of the spin-spin Green function becomes:
\begin{equation}
\label{eq:SpinGF}
\left<\left< X_{\vec{q}}^{\sigma \bar\sigma} \left| \right. X_{\vec{q}}^{\bar\sigma \sigma} \right>\right>_\omega 
= \frac{A_{\vec{q}}(\omega)}{\omega^2 - \omega_{\vec{q}}^2},
\end{equation}
where
\begin{eqnarray}
A_{\vec{q}}(\omega) &=& \frac{1}{N} \sum\limits_{\vec{k}} 
\left[ \left( - 2 t_{\vec{k}} + \frac{1-x}{2} \frac{\tilde{t}_k^2}{E_{ct}} \right) \left( K_{\vec{k}} - K_{\vec{q}-\vec{k}} \right) \right. \nonumber\\
&+& \left. 4 J_{\vec{k}} \left( C_{\vec{q}-\vec{k}} - C_{\vec{k}} \right) \right],
\end{eqnarray}
and magnetic excitations spectrum $\omega_{\vec{q}}$ represented by the Eq.~(26) of paper~\cite{valkov2005}.

The following results were obtained by self-consistent calculation of
the chemical potential $\mu$,
the spin-spin correlation functions~(\ref{eq:C}) using Green function~(\ref{eq:SpinGF}),
and the kinematic correlation functions~(\ref{eq:K}) using Green function~(\ref{eq:D_H})
with the self-energy~(\ref{eq:Sigma}).

\section{Results for \lowercase{p}-type cuprates \label{section:ptype}}

For LSCO the LDA+GTB calculated parameters are (in eV): 
$t=0.93, t'=-0.12, t''=0.15, J=0.295, J'=0.003, J''=0.007$.
All figures below are in electron representation.

\begin{figure}
\includegraphics[width=\linewidth]{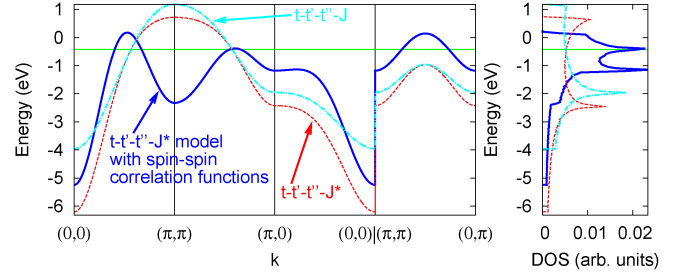}
\caption{(color online) The quasiparticle dispersion (on the left) and 
the density of states (DOS, on the right) in the paramagnetic phase 
of p-type cuprate with $x=0.16$. 
The position of the chemical potential is denoted by the horizontal (green) line. 
Results within the Hubbard-I approximation are shown by 
dashed (red) and dash-dotted (cyan) curves 
for the model with ($t-t'-t''-J^*$ model) 
and without ($t-t'-t''-J$ model) three-site correlated hoppings, respectively.
Bold solid (blue) curves represent the results for the 
$t-t'-t''-J^*$ model with the short range magnetic order.}
\label{fig:models}
\end {figure}

First of all we would like to stress the important effects caused 
by the three-site correlated hoppings $H_3$ and the renormalizations 
due to the short range magnetic order. 
Previously, the importance of the three-site correlated hoppings in the normal 
and superconducting phases has been demonstrated in Refs.~\cite{valkov2005,valkov2002,korshunov2004}.
In Fig.~\ref{fig:models} we present our results for 16\% hole doping within different approximations.
Evidently, introduction of three-site interaction terms results in the change 
of the position of the top of the valence band. 
Therefore, this will become important at small $x$. 
In AFM phase of the $t-J$ model there is a symmetry around $(\pi/2,\pi/2)$ point. 
In the paramagnetic phase this symmetry is absent. 
Due to the scattering on the short range magnetic fluctuations with AFM wave vector $\vec{Q}=(\pi,\pi)$
the states near the $(\pi,\pi)$ point are pushed below the Fermi level (see Fig.\ref{fig:models}),
thus totally changing the shape of the FS. In other words, the short range magnetic order
``tries'' to restore the symmetry around $(\pi/2,\pi/2)$ point. 
In our calculations the short range magnetic fluctuations are taken 
into account via the spin-spin correlation functions~(\ref{eq:C}).

\begin{figure}
\includegraphics[width=1\linewidth]{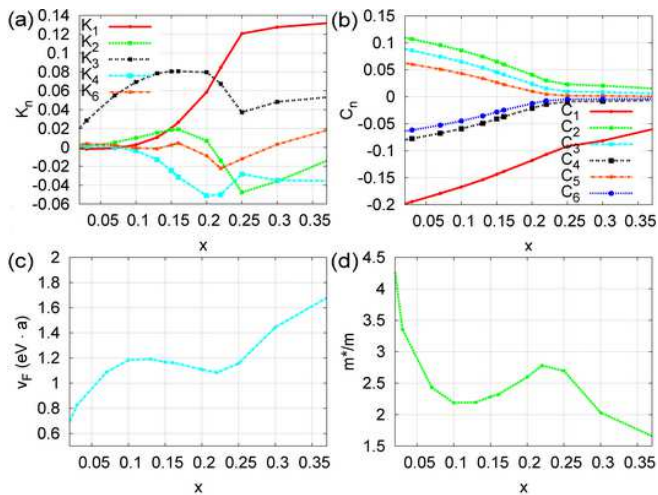}
\caption{(color online) Doping dependent evolution of the kinematic (a) 
and spin-spin (b) correlation functions within the $t-t'-t''-J^*$ model 
for p-type cuprates.
Index $n$ enumerates real space vectors connecting neighboring sites: 
$n=1$ for nearest-neighbors, $n=2$ for the next nearest neighbors, and so on.
In (c) and (d) the doping dependence of the nodal Fermi velocity 
(in units of eV$\cdot a$ with $a$ being a lattice constant) 
and the effective mass are shown.
}
\label{fig:tJ3pC}
\end {figure}

Our results for the doping dependence of the kinematic and 
spin-spin correlation functions are shown in Fig.~\ref{fig:tJ3pC}.
Note, the kinematic correlation functions $K_n$ possess a very nontrivial doping dependence.
For low concentrations, $x<0.15$, due to the strong magnetic correlations the hoppings to
the nearest and to the next-nearest neighbors are suppressed leading to the small values of 
$K_1$ and $K_2$, while $K_3$ is not suppressed.
Upon increase of the doping concentration above $x \approx 0.15$, 
magnetic correlations decrease considerably and nearest-neighbor 
kinematic correlation function $K_1$ increase.
Next major change sets at $x \approx 0.23$ when the system 
possesses almost Fermi liquid behavior:
$K_1$ becomes largest of all $K_n$'s, while the magnetic 
correlation functions $C_n$ and the kinematic correlation 
function $K_3$ are strongly suppressed.

\begin{figure}
\includegraphics[width=\linewidth]{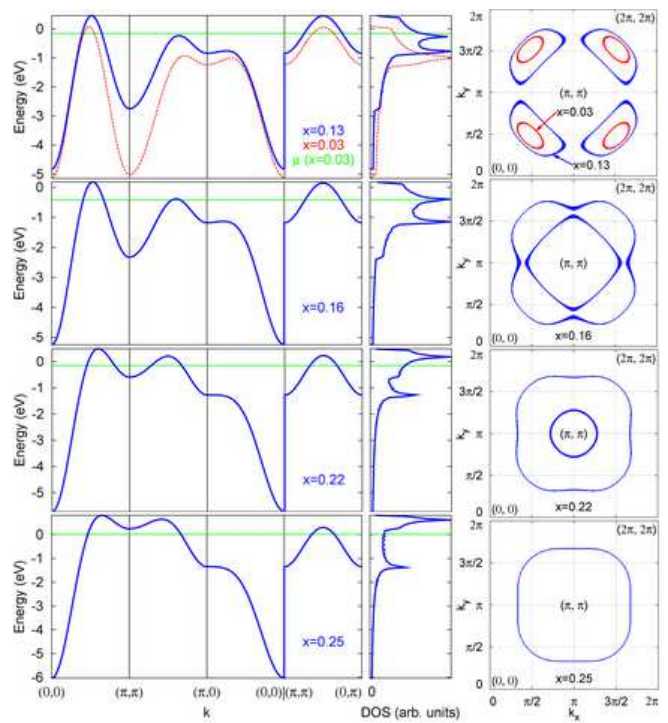}
\caption{(color online) Band structure (on the left), 
density of states (in the middle), and Fermi surface (on the right)
evolution with doping concentration $x$ 
within the $t-t'-t''-J^*$ model for p-type cuprates.}
\label{fig:tJ3pBands}
\end {figure}

So, we can clearly define two points of the crossover, namely $x \approx 0.15$ and $x \approx 0.23$.
The system behavior is quite different on the different sides of these points,
although there is no phase transition with symmetry breaking occurs.
To understand the nature of these crossovers we consider the FS evolution with doping concentration,
presented in Fig.\ref{fig:tJ3pBands}.
At low $x$ the FS has the form of the hole pockets centered around $(\pm \pi/2,\pm \pi/2)$ point.
Then these pockets enlarge and at $x=0.15$ all of them merge together,
forming the two FS contours. Up to $x=0.23$ the FS topologically equivalent
to the two concentric circles with the central one shrinking toward $(0,0)$ point.
For $x=0.23$ the central FS contour shrinks to the single point and vanishes,
leaving one large hole-type FS.

Apparently, the topology of the FS changes drastically upon doping. 
In particular, it happens at $x=0.15$ and at $x=0.23$.
For the first time the ``electronic transition'' accompanying the change in the FS topology, 
or the so-called Lifshitz transition, was described in Ref.~\cite{lifshitz1960}.
Now such transitions referred as a quantum phase transitions 
with a co-dimension$=1$ (see e.g. paper~\cite{volovik2006}).
Note, when the FS topology changes at quantum critical concentrations $x_1=0.15$ and at $x_2=0.23$
the density of states at the Fermi level also exhibit significant modifications.
This results in the different behavior of the kinematic and magnetic correlation functions
on the different sides of these crossover points.
And the changes in the density of states at the Fermi level will also 
result in the significant changes of such observable physical quantities as 
the resistivity and the specific heat.

Also, from the obtained quasiparticle dispersion 
we calculated the doping dependence of the nodal Fermi velocity $v_F$ and 
the effective mass $m^*/m$ (see Fig.~\ref{fig:tJ3pC}(c) and (d)).
Nodal Fermi velocity does not show steep variations 
with increase of the doping concentration in agreement 
with the ARPES experiments \cite{zhou2003,kordyuk2005}. 
Effective mass $m^*$ increase with decreasing $x$ and 
reveals tendency to the localization in the vicinity of 
the metal-insulator transition. But this increase is not 
very large and overall $m^*/m$ doping dependence agrees 
quite well with the experimentally observed one~\cite{padilla2005}.
Note, the non-monotonic doping dependence of both these quantities 
reflects the presence of the critical concentrations $x_1$ and $x_2$.

\begin{figure}
\includegraphics[width=1\linewidth]{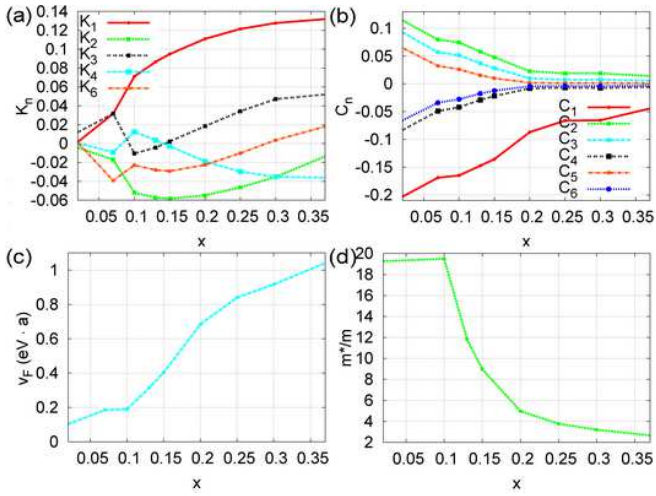}
\caption{(color online) The same as in Fig.\ref{fig:tJ3pC}, 
but within the model for p-type cuprates without three-site correlated hoppings ($t-t'-t''-J$ model).}
\label{fig:tJpC}
\end {figure}

\begin{figure}
\includegraphics[width=\linewidth]{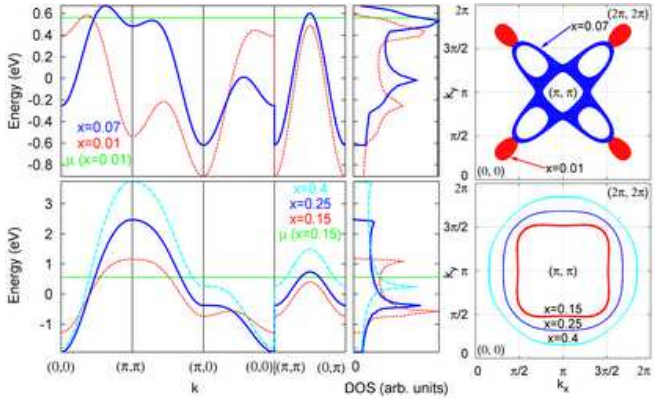}
\caption{(color online) The same as in Fig.\ref{fig:tJ3pBands}, 
but within the model for p-type cuprates without three-site correlated hoppings ($t-t'-t''-J$ model).}
\label{fig:tJpBands}
\end {figure}

To analyze the effect of the three-site hopping term $H_3$
we also calculated the doping dependence of the band structure and FS
within the $t-t'-t''-J$ model. 
The behavior of the kinematic and the spin-spin correlation functions, 
presented in Fig.~\ref{fig:tJpC}, is quite different from that 
of the $t-t'-t''-J^*$ model. 
There is only one quantum critical point at $x \approx 0.08$ and
the effective mass becomes very large for $x$ approaching zero.
For $x>0.1$ the evolution of the FS and density of states 
near the Fermi level is smooth, without significant changes (see Fig.\ref{fig:tJpBands}).
Most part of the difference to the $t-t'-t''-J^*$ model stems from the
role of $H_3$ in the energy of states near the $(\pi,\pi)$ point,
thus determining the topology of the FS and the physics at low doping 
concentrations (see Fig.\ref{fig:models}).

\section{Results for \lowercase{n}-type cuprates \label{section:ntype}}

Now let us consider n-type cuprates within the $t-t'-t''-J^*$ model.
For NCCO the LDA+GTB calculated parameters are (in eV): 
$t=-0.50, t'=0.02, t''=-0.07, J=0.195, J'=0.001, J''=0.004$.

\begin{figure}
\includegraphics[width=1\linewidth]{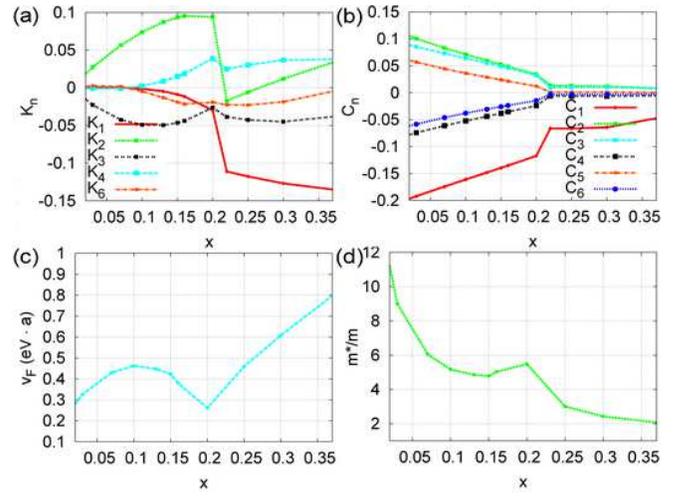}
\caption{(color online) The same as in Fig.\ref{fig:tJ3pC}, but within the $t-t'-t''-J^*$ model 
for n-type cuprates.}
\label{fig:tJ3nC}
\end {figure}

\begin{figure}
\includegraphics[width=\linewidth]{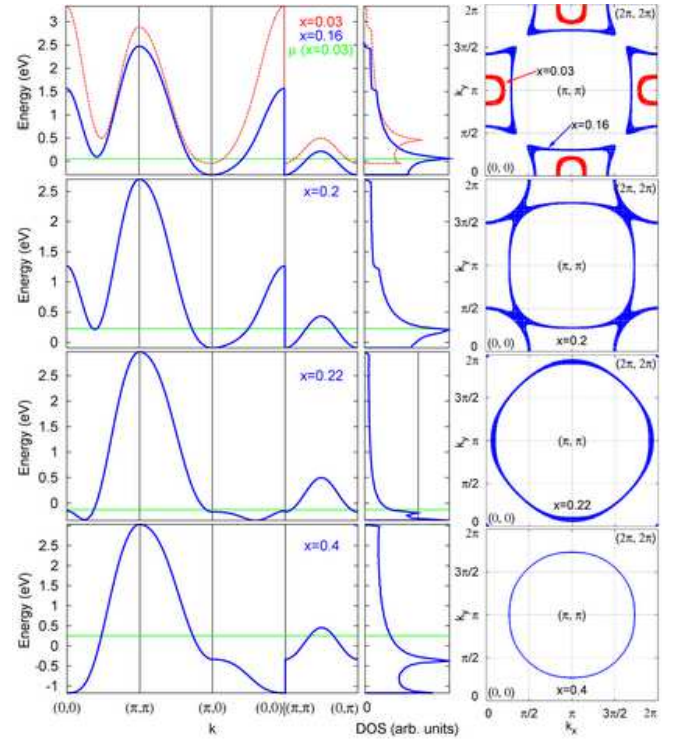}
\caption{(color online) The same as in Fig.\ref{fig:tJ3pBands}, but within the $t-t'-t''-J^*$ model 
for n-type cuprates.}
\label{fig:tJ3nBands}
\end {figure}

The obtained doping dependence of the kinematic and magnetic correlation functions 
presented in Fig.~\ref{fig:tJ3nC}. There is only one crossover point at $x \approx 0.2$.
Also, in contrast to the p-type results, the most important kinematic correlation function
on the left of this point is $K_2$, rather than $K_3$.
For $x>0.2$ the system demonstrates Fermi liquid-like behavior with
magnitude of kinematic correlation function decreasing with the distance,
and small values of the magnetic correlations.

The role of the short range magnetic order and three-site hopping terms
in n-type cuprates is similar to that of p-type. In particular, due to the scattering 
on the magnetic excitations the states near the $(\pi,\pi)$ point pushed
{\it above} the Fermi level, and the local symmetry around the $(\pi/2,\pi/2)$ 
points is restored, reminding of the short-range AFM fluctuations (see Fig.~\ref{fig:tJ3nBands}).

Instead of hole pockets around the $(\pm \pi/2,\pm \pi/2)$ point in p-type,
here at low $x$ the electron pockets around $(\pm \pi,0)$ and $(0, \pm \pi)$ points 
are present. Upon increase of the doping concentration these pockets become larger 
and merge together at $x=0.2$. For higher concentrations the FS appear to be
a large hole-like one, shrinking toward $(\pi,\pi)$ point.
Therefore, no other changes in the FS topology other than at $x=0.2$ are present.
Referring to the same arguments as in previous section, we claim that in our approach
there is only one quantum critical point at $x_n=0.2$ in the n-type cuprates.
Th non-monotonic change of the effective mass and the nodal Fermi velocity 
is also present at this concentration, as evident from Fig.~\ref{fig:tJ3nC}(c) and (d).

\section{Discussion and summary \label{section:discussion}}

To summarize, we have investigated the doping-dependent evolution of the low-energy excitations 
for p- and n-type High-$T_c$ cuprates in the regime of strong electron correlations within 
the sequentially derived effective model with the {\it ab initio} parameters.
We show that due to the changes of the Fermi surface topology with doping
the system exhibits drastic change of the low-energy physics. 
Namely, for p-type cuprates there exist two critical concentrations, 
$x_1 \approx 0.15$ and $x_2 \approx 0.23$. 
Along the different sides of these concentrations the behavior of
the density of states near the Fermi level, of the kinematic and 
magnetic correlation functions, of the effective mass 
and the nodal Fermi velocity, is drastically different.
This let us speak about crossover, or, taking into account the 
accompanying FS topology changes, about quantum phase transitions 
at these quantum critical concentrations.

For n-type cuprates due to the specific FS topology 
we obtain only one quantum critical concentration, $x_n \approx 0.2$.

First of all, we would like to comment on the approximations made in this work.
Since we use the perturbation theory with hopping $t$ and exchange $J$ as small parameters,
appropriate for the strongly correlated regime, the real part of the corrections 
to the results obtained will be small to the extent of smallness of the higher powers of $t/E_{ct}$ and $J/E_{ct}$.
This will result in the small changes of the band dispersion 
and in the fine details of the FS, not changing anything qualitatively.

More concerns give the imaginary part of the neglected corrections to the 
strength operator $\hat{P}(\vec{k},E)$ and to the 
self-energy $\hat{\Sigma}(\vec{k},E)$ through $\tilde{L}_{\vec{k}}^{\alpha (irr)}$.
Application of the equation of motion decoupling method to the Hubbard model 
with finite quasiparticle lifetime~\cite{plakida2006} 
reveals that the results of the mean-field-like approximation is qualitatively correct.
Quantitatively, at low doping the imaginary part of the self-energy leads 
to the hiding of the FS portions above the antiferromagnetic Brillouin 
zone ($(\pi,0)-(0,\pi)$ line). This results in Fermi arc rather than hole 
pockets at $x<x_{opt}$ (see Fig.~\ref{fig:tJ3pBands}).
Also, we can compare our results to the numerical methods, namely, 
to the exact diagonalization studies \cite{tohyama2004}.
The quasiparticle dispersion of the $t-t'-J$ model 
in Ref.~\cite{tohyama2004} can be considered as consisting of two bands. 
For p-type cuprates intensities of the spectral peaks corresponding 
to the band situated mostly above the Fermi level (in electron representation) 
are very low. This band is often called a ``shadow'' band and appears due 
to the scattering on the short range AFM fluctuations. Notably, our band dispersion from 
Figs.~\ref{fig:tJ3pBands}~and~\ref{fig:tJ3nBands} reproduce very well 
the shape of the other, ``non-shadow'', band. It is this band where 
the most part of the spectral weight is residing, thus determining most 
of the low-energy properties, except for such subtle effects as a so-called ``kink'' 
in dispersion \cite{zhou2003}.

Also, all renormalizations not included in consideration will change
the values of the critical concentrations $x_1$, $x_2$, and $x_n$. 
Comparing with the results of a more rigorous theory in paper \cite{plakida2006}, 
we expect these values to decrease.

Thus we conclude that our theory captures the most important part 
of the low-energy physics within the considered (and justified for cuprates) model.
This claim is supported by the qualitative agreement with the critical concentrations 
of crossover observed in the transport experiments \cite{ando2004RH,ando2004}
and in the optical experiment \cite{gedik2005,bontemps2004,carbone2006},
and even quantitative agreement of the doping dependence of the nodal Fermi velocity 
and of the effective mass \cite{zhou2003,kordyuk2005,padilla2005}.
Although we use a simple mean-field theory, though strong-coupled, 
the agreement with the experiments is not surprising 
since we included all necessary for High-$T_c$ copper oxides ingredients.
Namely, the short range magnetic order and three-site correlated hoppings.
Former is the intrinsic property of the cuprates exhibiting long range AFM order at low $x$,
while latter results from the sequential derivation of the low-energy effective model.

\begin{acknowledgments}
Authors would like to thank D.M. Dzebisashvili and V.V. Val'kov for very helpful discussions,
I. Eremin, P. Fulde, N.M. Plakida, A.V. Sherman, and V.Yu. Yushankhai for useful comments. 
This work was supported by INTAS (YS Grant 05-109-4891), 
Siberian Branch of RAS (Lavrent'yev Contest for Young Scientists), 
RFBR (Grants 06-02-16100 and 06-02-90537), 
Program of Physical Branch of RAS ``Strongly correlated electron systems'', 
and Joint Integration Program of Siberian and Ural Branches of RAS N.74.
\end{acknowledgments}

\end{document}